# Spintronics, from Giant Magnetoresistance to magnetic Skyrmions and Topological Insulators

Albert Fert[1] and Frédéric Nguyen Van Dau[1]

[1]Unité Mixte de Physique, CNRS, Thales, Univ. Paris-Sud, Université Paris-Saclay, 91767, Palaiseau, France

**This article aims at giving a general presentation of spintronics, an important field of research developing today along many new directions in physics of condensed matter. We tried to present simply the physical phenomena involved in spintronics – no equations but many schematics. We also described the applications of spintronics, those of today and those expected to have an important impact on the next developments of the information and communication technologies. This article was published in the Comptes Rendus Physique of the French Academy of Sciences (https://doi.org/10.1016/j.crhy.2019.05.020).**

## 1  Introduction

Spintronics is generally defined as a new type of electronics manipulating electrons by acting not only on the charge of the electrons but also on their spin. Its development started with the discovery [1], [2] of the giant magnetoresistance (GMR) of magnetic multilayers in 1988. The new concepts on the manipulation of spin-polarized currents introduced by the GMR and, in addition, the potential of GMR for applications boosted the development of an intense activity in a field of research that was later called spintronics. In the experiments of these first times of spintronics, the spin-polarized currents were generated by using the influence of the orientation of the spin on the transport properties of the electrons in ferromagnetic conductors. This influence, first suggested by Mott [3], had been experimentally demonstrated and theoretically described in early works [4]–[6] more than ten years before the discovery of the GMR. This type of production of spin-polarized currents by magnetic materials was used in the "classical spintronics" of the first decade after GMR discovery. Major events of this time were the discoveries of the Tunneling Magnetoresistance (TMR) and Spin Transfer Torque. It was also the time of the introduction of important concepts as those of spin accumulation and pure spin current (current of spin without current of charge). More recently, it became possible to produce spin polarized currents and pure spin currents without magnetic materials by using the spin-orbit interactions in non-magnetic materials, in a field that has been called spin-orbitronics. Today spintronics is expanding in many directions. The promising new axes that we will discuss are spintronics with topological systems (as the interfaces states of topological insulators), and spintronics with magnetic skyrmions.

## 2  Roots of spintronics

### 2.1  Spin dependent conduction in ferromagnetic metals
The dependence of the electrical conduction in ferromagnetic metals on the orientation of the electron spin with respect to the direction of the magnetization can be understood from the typical band structure of a ferromagnetic metal shown in Fig.1a. As there is an energy splitting between the "majority spin" and "minority spin" bands (spin up and spin down in the usual notation), the electrons at the Fermi level, which carry the electrical current, are in different states and exhibit different conduction properties for opposite spin directions. Such a dependence of the conduction on the spin orientation was proposed by Mott [3] in 1936 to explain some features of the resistivity of ferromagnetic metals at the Curie temperature. However, in 1966, at the beginning of the Ph.D. thesis of one of the authors (AF), the subject was still almost completely unexplored. The experiments of this thesis were resistivity measurements on Ni or Fe doped with different types of impurities and, from the analysis of the temperature dependence of the resistivity and also, mainly, from experiments on ternary alloys described in the next paragraph, it could be shown that the resistivities of the two channels can be very different in metals doped with specific impurities presenting a strongly spin dependent scattering cross-section [4]–[6] . In Fig.1b, we show the example of the spin up and spin down resistivities of nickel doped with 1% of different impurities. It can be seen that the ratio α of the spin down resistivity to the spin up one can be as large as 20 for Co impurities or, as well, smaller than one for Cr or V impurities. This was consistent with the theoretical models developed at this time by Jacques Friedel and Ian Campbell for the electronic structures of impurities in metals. The two current conduction was rapidly confirmed in other groups and, for example, extended to Co-based and Fe-based alloys [7], [8].

## 2.2 Experiments on Ni-based ternary alloys anticipating the concept of GMR

Let us consider resistivity measurements on ferromagnetic metals doped with two types of impurities [4]–[6], [8], [9]. Suppose, as illustrated by Fig.2, a ternary alloy of nickel doped with 0.5% of impurities A (Co for example) which scatter strongly the electrons of the spin down channel and, in the same ternary alloy, 0.5% of impurities B (Rh for example) which scatter strongly the spin up electrons. In such a ternary alloy Ni(Co, Rh), that we call of type #1, the electrons of both channels are strongly scattered either by Co in one of the channels or by Rh in the other, so that there is no shorting by one of the channels as when Ni is doped only with either 1% of Co or 1% of Rh, and the resistivity is strongly enhanced. For a total impurity concentration of 1% and variable proportions x and (1-x) of the Co and Rh concentrations, as illustrated in Fig.2a, the resistivity is definitely enhanced by a mixed doping of Co and Rh. On the contrary, there is no such enhancement in ternary alloys of type #2 doped by impurities A and B (Co and Au in Fig.2b) scattering strongly the electrons in the same channel and leaving the second channel open.

The concept of GMR is the replacement of the impurities A and B of the ternary alloy by the successive layers A and B of the same magnetic metal in a multilayer. If the magnetizations of the layers A and B are antiparallel, this corresponds to the situation of strong scattering in both channels for ternary alloys of type #1 (Fig.2a). In contrast, the configuration with parallel magnetizations corresponds to the situation with a "free channel" in all the layers as in ternary alloys of type #2. What is new with respect to the ternary alloys is the possibility of switching between high and low resistivity states by simply changing the relative orientation of the magnetizations of layers A and B from antiparallel to parallel. However, the transport equations tell us that the relative orientation of layers A and B can be felt by the electrons only if their distance is smaller than the electron mean free path, that is, practically, if they are spaced by only a few nm. Unfortunately, in the seventies, it was not technically possible to make multilayers with layers as thin as a few nm and the discovery of the GMR waited until the development of sophisticated deposition techniques.

## 3 The discovery of GMR (Giant Magnetoresistance) and applications of GMR

In the mid-eighties, with the development of techniques like the Molecular Beam Epitaxy (MBE), it became possible to fabricate multilayers composed of very thin individual layers and one could consider trying to extend the experiments on ternary alloys to multilayers. In addition, in 1986, Peter Grünberg and co-workers [10] discovered the existence of antiferromagnetic interlayer exchange couplings in Fe/Cr multilayers. Fe/Cr appeared as a magnetic multi-layered system in which it was possible to switch the relative orientation of the magnetization in adjacent magnetic layers from antiparallel to parallel by applying a magnetic field. In collaboration with a group of at the Thomson-CSF company, the authors (AF and FVD) started the fabrication and investigation of Fe/Cr multilayers. This led in 1988 to the discovery [1] of very large magnetoresistance effects (Fig.3a) that was called GMR. Peter Grünberg and co-workers at Jülich [2] obtained effects of the same type in Fe/Cr/Fe trilayers practically at the same time. The interpretation of the GMR is similar to that described above for the ternary alloys and is illustrated by Fig.3b. The first classical model of the GMR was published in 1989 by Camley and Barnas [11] and Levy and co-workers worked out the first quantum model [12] in 1991.

GMR attracted rapidly attention for its fundamental interest as well as for the many possibilities of applications. The research on magnetic multilayers and GMR developed rapidly. On the experimental side, two important results were published in 1990. Parkin et al [13] demonstrated the existence of GMR in multilayers made by the simpler and faster technique of sputtering and found the oscillatory behaviour of the GMR due to the oscillations of the interlayer exchange as a function of the thickness of the nonmagnetic layers. Shinjo and Yamamoto [14], as well as Dupas et al [15], demonstrated that GMR effects can be found in multilayers without antiferromagnetic interlayer coupling but composed of magnetic layers of different coercivities. Another result, in 1991, was the observation of large and oscillatory GMR effects in Co/Cu, which became an archetypical GMR system [16], [17]. Also in 1991, Dieny et al [18], reported the first observation of GMR in spin-valves, i.e. trilayered structures in which the magnetization of one of the two magnetic layers is pinned by coupling with an antiferromagnetic layer while the magnetization of the second one is free. The magnetization of the free layer can be reversed by very small magnetic fields, so that the concept is now used in many devices [19].

GMR had quickly various applications, first several types of magnetic sensors used today in the car industry. Then, in 1996, came the application to the read heads of the hard disk drives (Fig. 3c,d) which is certainly the most economically important [20], [21]. The GMR and similar magnetoresistance effects called TMR and CPP-GMR (see next section) provided sensitive read techniques and led to an increase of the areal recording density by more than two orders of magnitude (from $\approx 1$ to $\approx 1000$ Gbit/in2 today). This increase opened the way both to highly improved storage capacities (up to several terabytes) for video recording or backup and to smaller HDD sizes (down to .85-inch disk diameter) for mobile appliances like ultra-light laptops or portable multimedia players. GMR sensors are also used in many other types of application, for example the compass that we have today in some smartphones and series of bio-medical devices [22].

## 4  Classical spintronics

After the early years spent to extensively study GMR, spintronics became a broader research domain after the successive discoveries of several physical effects, all relying on the role of spin on electronic transport. In this section, we review the most important steps, in this classical spintronics phase, and their impact on applications.

### 4.1  CPP-GMR, spin accumulation and pure spin currents

As the first magnetic multilayers were made of metallic magnetic or non-magnetic materials, the first GMR experiments were carried out in a straight forward experimental configuration where current is flowing along the plane of the layers (CIP geometry). From 1991, it has been possible to investigate GMR when the current is flowing perpendicularly to the layers (CPP Geometry), first by using superconducting electrodes [23], then by electrodepositing multilayers into the nanopores of a membrane [24]. In the CPP geometry, GMR was found to be substantially higher than in the CIP geometry, and to subsist up to much larger layer thicknesses. The Valet-Fert model [25] explains that, owing to spin accumulations occurring when the current crosses interfaces between layers, the length scale in the CPP geometry becomes the long spin diffusion length (SDL) instead of the short mean free path limiting the CIP GMR to multilayers with very thin layers. The CPP-GMR studies have clearly revealed the spin accumulation effects, which govern the propagation of a spin-polarized current through successive layers of magnetic and non-magnetic materials and have played an important role in the subsequent developments of spintronics.

The physics of spin accumulation occurring when an electron flux crosses an interface between a ferromagnetic and a non-magnetic material is explained in Fig. 4a for a simple situation (single interface, no interface resistance, no band bending and single polarity). To summarize, there is a broad zone of spin accumulation which extends on both sides of an F/N interface with an exponential decreases at the scale of the spin diffusion length of each material. The progressive transition between spin-polarized current and non-polarized current occurs by spin-flips in this zone.

The physics of spin accumulation and spin-polarized currents can be described by new types of transport equations in which the electrical potential is replaced by a spin-and position-dependent electrochemical potential [25]–[27]. These equations can also be extended to take into account band bending and high current density effects [28]. These equations can be applied to the general situation of multi-contact systems with interplay of the spin accumulation occurring around the different interfaces [29], [30]. A standard structure for this situation is illustrated by Fig.4b. A spin polarized current is injected from the ferromagnetic contact F1 into the non-magnetic channel N and flows to the left side of N. The blue and red curves represent the spin up and spin down electro-chemical potentials associated with the spin accumulation at the F1/N interface. The opposite gradients of these electro-chemical potentials (pseudo electric fields) on the right side of N generate currents of spin up and spin down electrons in opposite directions. This current distribution is described as a pure spin current which does not carry any charge but only a flux of spin momentum.

### 4.2 Tunnel Magnetoresistance (TMR)

In a magnetic tunnel junction (MTJ) where an ultrathin tunnel barrier is separating two ferromagnetic electrodes, TMR refers to the change of junction resistance, depending on the magnetic configuration of the electrodes (parallel or antiparallel magnetizations). Early observations of TMR were reported in 1975 by Jullière [31] and in 1982 by Maekawa and Gäfvert [32]. These pioneering works did not follow immediately, probably due to the difficulty to reproduce them. In 1995, large (~20%) and reproducible TMR were reported by Moodera's [33] and Miyasaki's [34] groups on MTJs using a tunnel barrier made of an oxidized ultrathin Al layer.

These original contributions have then triggered an intense activity around the development of MTJs. One important step has been the transition from MTJs using amorphous tunnel barriers (mainly oxidized Al) to single-crystal MTJs, in particular using MgO barrier. Whereas the first results obtained with MgO-based MTJ were only slightly larger than with alumina barriers [35], a breakthrough was achieved in 2004 with the first report of large TMR ratios (~200% at room temperature, Fig. 4c,d) in epitaxial MTJs at Tsukuba [36] and IBM [37]. Continuous progress have been achieved since 2004 with MgO-based MTJs and TMR ratios have increased above 600% [38].

The origin of large TMR in MTJ having a single-crystal barrier such as MgO has been attributed to symmetry selection [39]–[42]. At a high quality interface between a magnetic electrode and a tunnel barrier, an evanescent wave function of a given symmetry is connected to the Bloch functions of the same symmetry and spin direction at the Fermi level of the electrodes. In (001)MgO barriers, one of the symmetry corresponding to highly spin-polarized bands has a much slower decay, which results in a higher TMR. Other symmetry rules have been observed in barriers made of different materials, leading to different properties such as negative effective spin polarization in $SrTiO_3$-based MTJ [43].

Other research directions have been followed in order to further increase TMR ratios such as the replacement of conventional ferromagnetic electrodes by 100% spin polarized half metals made of oxide materials [44] or Heussler alloys [45], or using ferromagnetic barriers as a spin filter [46].

From an applicative point of view, TMR have been introduced by the industry in read-heads of hard disk drives (HDD) in 2006, mainly because it was bringing improvements in terms of signal to noise ratio. This has allowed sustaining the continuous increase of areal densities of stored information and, at the same time, the cost reduction for this storage. Since the introduction of GMR in HDD read heads in 1997, the cost per GByte stored in a desktop have dropped by more than three orders of magnitude, from $100 to less than 3 cents. This has been a clear driver in the emergence of the big data era [47]. Another domain of applications that was impacted by TMR was the emergence of magnetic random access memories (MRAM). Whereas the concept of MRAM was already introduced by Honeywell before the discovery of GMR [48], [49], its development has only been triggered by the late 1990's using TMR. Indeed, the natural configuration of MTJ building block, with current flowing perpendicular to layers, was particularly well suited to the development of compact architectures such as cross bars. However, in the first generation of TMR-based MRAM developed by the mid-2000's, writing information was achieved through magnetic fields generated by "word" and "bit" lines, which proved to be expensive in terms of power consumption and also lead to problems of cross-talk. This was solved by the introduction of spin transfer torque writing [21]. Indeed, it was realized at that time that the impossibility to locally generate a magnetic field was a real hurdle (in word and bit lines, only the magnetic field generated at the cross point is useful, the rest being wasted energy). This has triggered the subsequent research toward an electrical control of magnetization, using spin transfer torque (see 4.3), multiferroics (combining ferroelectric and ferromagnetic properties) [50][51], up to the recent Intel's MESO device concept [84]. In this concept, an input voltage pulse is used to switch a multiferroic/ferromagnetic bilayer and the magnetization of the magnetization direction of the ferromagnetic layer is converted into an output voltage by the Edelstein effect of a topological bi-dimensional electron gas (see 5.2).

**4.3  Spin Transfer Torque (STT)**

STT relates to the manipulation of the magnetic moment of a ferromagnetic element without applying any magnetic field but only by transferring spin angular momentum by a spin polarized current. The concept was first introduced theoretically by Slonczewski [52] and Berger [53] and is illustrated in Fig. 5a. The injection of a spin polarized current into a magnetic element leads to the transfer of the transverse component of the spin current, which creates a torque acting on the magnetization of the element. This spin transfer torque can be used to either switch the magnetization (Fig. 5b,d) or to create a steady-state gyration regime (Fig. 5c,e). In the latter case, GMR or TMR can be used to translate this magnetization gyration into an oscillating voltage in the microwave frequency range, leading to the concept of spin transfer oscillator.

The first experimental evidence of spin transfer were obtained using either point contacts [54], [55] or on pillar-shaped elements [56], [57]. Whereas the first experimental investigations of spin transfer were carried out on uniformly magnetized elements, the use of non uniform configurations such as magnetic vortices has proved to open interesting perspectives. In particular, it has allowed to strongly increase the power emission and decrease the linewidth of spin transfer oscillators [58]. Spin transfer torque has also be used to move domain walls in nanostructures [59] leading to the concept of racetrack memory.

As far as applications are concerned, the appearance of STT has undoubtedly boosted the development of MRAM, which are now called STT-RAM. Indeed, it allows to significantly reduce the power consumption associated with the writing function while also improving reliability with respect to cross talk, in particular. Since the first demonstrations of STT-RAM in the mid-2000's [60] this new product has often been presented as a potential universal electronic memory. It has indeed arguments to compete with the three main types of electronic memories: DRAM and SRAM to which it will bring non volatility (to compete with SRAM, one must also face a very hard speed challenge) and embedded Flash to which it will bring a better write endurance and a simpler fabrication process. Since the introduction of the first STT-RAM product on the market in 2012 by Everspin [61], several companies have been active in the industrialization. During the last two years, several major players have announced the start of mass production [62].

## 5 Spin-orbitronics

In classical spintronics, during the first decade of its development, magnetic materials were used to spin-polarize the current. An important change, a few years after 2000, came from the possibility of creating spin-polarized current and pure spin currents by using the spin-orbit coupling (SOC, relativistic correction to the equations of quantum physics coupling spin and momentum). In magnetic materials, the SOC is known to be at the origin of properties as the magnetocrystaline anisotropy, the Anomalous Hall Effect (AHE), the Anisotropic Magnetoresistance (AMR). In nonmagnetic conductors, the SOC generates the Spin Hall Effect (SHE), which appeared to be very efficient to convert charge currents into spin currents and vice-versa, which is the basic operation in any spintronic devices. Today, the SHE of heavy metals having a large SOC is used in several spintronic devices. We will present successively the SHE, then the similar but more efficient effects obtained by exploiting SOC effects in bi-dimensional electron gas at the surface of topological insulators or at Rashba interfaces, and finally the fast development of the physics of the topological magnetic solitons called skyrmions.

### 5.1 Spin Hall Effect (SHE)
For currents flowing in non-magnetic materials, the SHE results from the SOC-induced deflection of electrons of opposite spins in opposite direction, as represented in Fig.6a with spin up and spin down electrons deflected respectively to the right and left of the conductor. This leads to opposite spin accumulations on opposite edges, for example spin up accumulated and spin down depleted on the left edge on the figure. On each edge, due to the balance between the accumulation and depletion of opposite spins, there is no charge accumulation, as for the Ordinary Hall Effect, and therefore no voltage between the Hall contacts (Fig. 6a). The SHE has been predicted by Dyakonov and Perel [63] in 1971, theoretically revisited more recently by Hirsh [64] and the spin accumulation on the edges has been seen for the first time by Kerr optical measurements on GaAs in 2004 [65]. Valenzuela and Tinkham [66], [67] performed the first demonstration by transport measurements with the device shown in Fig. 6b. A current flowing from the ferromagnetic channel to the nonmagnetic channel on the left (from A to B) generates a spin accumulation at their interface and this spin accumulation diffuses a pure spin current to the right of the nonmagnetic channel (as already represented in Fig. 4b, a spin current corresponds to opposite flows of opposite spins). With spin up and spin down electrons flowing in opposite directions, the SOC deflects both of them in the same direction of the transverse channel, which generates a nonzero charge current in this channel or, in the open circuit conditions of the figure, a nonzero voltage between C and D. Experimental results are in Fig. 6c.In term of conversion between charge and spin current, the type of experiment described in the preceding paragraph and Fig .6b-c represents a conversion of a pure spin current into a charge current (or voltage), what is generally described as an experiment of Inverse Spin Hall Effect (ISHE). Direct SHE corresponds to the opposite conversion in the typical situation of Fig.6d. As shown in the figure, the charge current in the heavy metal (Ta) deflects the blue and red spins downward and upward. At the top interface, the blue spins accumulate and this spin accumulation can diffuse into an adjacent conducting layer, injecting blue spins and sucking down red spins. In other terms, the SHE of the heavy metal injects a pure spin current into an adjacent layer, which corresponds to a conversion of the horizontal charge current into a vertical (pure) spin current.

Fig.6e represents a Spin-Orbit-Torque-MRAM (SOT-MRAM) exploiting the SHE of a Ta layer to inject a pure spin current into the bottom magnetic layer of a M-RAM memory and the Spin Transfer Torque induced by this injection of spin current switches its magnetization [68].

The SHE and ISHE are characterised by the spin Hall angle $\theta$, defined as the ratio between the SHE-induced transverse spin current density and the longitudinal charge current density, also equal to the ratio between the ISHE-induced transverse charge current density and the longitudinal spin current density. Actually, $\theta$ characterises the ratio between spin and charge currents inside an infinite material. If the material of Spin Hall angle $\theta$ is interfaced with other materials, the efficiency of the final conversion depends also on the interface transparency and properties of the other materials.

The larger SHE effects can be found in materials with heavy elements of large SOC [69]. The most used in spintronics are heavy metals such as Pt ($\theta \approx 0.05$ [70]), Ta ($\theta \approx 0.15$ [71]) or W $\theta \approx 0.3$ [72], [73]. In pure metals the SHE is generally an intrinsic property related to the effect of the SOC on the band structure (Berry curvature). Large Spin Hall angles can also be obtained in metals of small SOC doped with heavy nonmagnetic impurities, as, for example, Cu doped with Bi [74]. In this case, the mechanism of the SHE is the skew scattering or, alternatively, the scattering by side-jump, both types of mechanisms induced by the SOC of the heavy impurity scattering potential. Actually, the SHE and Spin Hall angle induced by heavy nonmagnetic impurities in Cu had been already measured in early experiments almost 40 years ago. At this time, the technologies did not exist to fabricate nanostructures such as that of Fig.6b to unbalance the deflections of opposite spin in opposite directions. However, in Cu doped with heavy nonmagnetic impurities, adding a tiny concentration of paramagnetic Mn impurities and polarizing their spin by a field was used to unbalance the spin up and spin down currents and single out the SHE induced by skew scattering by the nonmagnetic impurities [75].

## 5.2    Spintronics with topological insulators or Rashba interfaces

As shown in the preceding section, the spin-orbit coupling (SOC) can be harnessed for the conversion between charge and spin currents by using the SHE or ISHE of heavy metal layers. More efficient conversions between spin and charge by SOC can be obtained in the bi-dimensional (2D) electron gas (2DEGs) at the surface/interface of topological insulators (TI) or at Rashba interfaces. Such 2DEGs are characterized by Fermi contours with helical locking between the spin and momentum degrees of freedom. Figure 7a presents an artist view of the dispersion Dirac cone of TI surface or interface states as well as the helical spin configurations of the corresponding Fermi contours. The ARPES image of the Dirac cone of the 2D states at the interface between the TI $\alpha$-Sn cone and Ag is shown in Fig.7b. Rashba interface states also characterized by such locking between spin and momentum.

As depicted in Fig.7c, the current carried by helically spin-polarized 2D electrons at a TI surface or interface is automatically associated with a nonzero spin polarization along the in-plane direction transverse to the current, an effect predicted in 1990 by Edelstein [76]. The spin polarization of the 2DEG can diffuse into an adjacent conducting layer and inject a pure spin current into it, in the same way as the SHE injects a spin current into an adjacent layer (Fig.6d). The creation of a pure spin current by the spin polarization in the 2DEG can be described as the conversion of a 2D charge current Jc into a 3D spin current Js [77], [78]. This Edelstein Effect (EE) at TI or Rashba interfaces has been demonstrated by measuring directly the spin-polarization of the interface states [79] or by measuring the Spin Tranfer Torque induced by the injected spin into an adjacent magnetic layers [77].

The Inverse Edelstein Effect (IEE) correspond to as inverse conversion [80], [81]. As depicted in Fig.~~1~~7e-f, the injection of a 3D vertical spin current into the 2DEG at the interface between the TI $\alpha$-Sn and Ag induces a 2D charge current Ic in the 2DEG.

The spin/charge conversions by EE and IEE can be characterized by the conversion coefficient $q_{ICS}$ = Js /Jc (inverse of length) for the EE [77] and the IEE length1 $\lambda_{IEE}$ = Jc/ Js (length) for the IEE [80]. In Fig.7f, we present on example of spin to charge conversion by IEE in spin pumping experiments on the interface states of the TI $\alpha$-Sn. The ferromagnetic resonance of the NiFe top layer in the device of Fig.7e pumps (injects) vertically a pure spin current into the 2DEG at Ag/$\alpha$-Sn interface and we show in Fig.7f the peaks of charge current in the 2DEG detected at the resonance field (around 100 mT). The results in Fig. 7f lead to a conversion is characterized by $\lambda_{IEE}$ = 2.1 nm for the IEE conversion length, which corresponds to a conversion larger than with the ISHE of heavy metals by about one order of magnitude. Similar results (EE and IEE) have been obtained with 2DEGs of Rashba interfaces [78], [82]. Particularly efficient conversions ($\lambda_{IEE}$ = 6.4 nm) have been found with the Rashba 2DEG at the interface between the insulating oxides SrTiO3 and LaAlO3 [83]. Such efficient spin/charge conversions by 2DEGS are promising for the developments of new types of devices such as the TI-based MESO logic devices proposed by the company Intel [84].

### 5.3  Magnetic skyrmions

A magnetic skyrmion is a topological swirling configuration of the spins of neighbor atoms in a magnetic material [85]–[88]. In Figures 8 a-b, we show schematic representations of the spin configuration in a magnetic skyrmion standing in a thin film uniformly magnetized in the up direction outside the skyrmion. The magnetization inside the skyrmion rotates progressively with a fixed chirality from the up direction at one edge to down at the center and again up on the opposite edge. This spin configuration can be mapped on a unit sphere and it covers all the possible axes on the sphere. Fig.8a (Néel-like skyrmion) and Fig. 8b (Bloch skyrmion) correspond to different directions of the rotation. Experimental images of skyrmions are shown in Fig.8c and 8d. A topological number S (or skyrmion number) characterizes the winding of the normalized local magnetization, **m**, which, in the two-dimensional limit, can be written as

$$S = \frac{1}{4\pi} \int \boldsymbol{m} \cdot (\partial_x \boldsymbol{m} \times \partial_y \boldsymbol{m}) \mathrm{d}x \mathrm{d}y = \pm 1, \quad \text{(Eq. 1)}$$

and equals + or – 1 for the configuration types of the figure. This peculiar topology of skyrmions gives rise to a topological protection of the structure: the spin configuration cannot be twisted continuously to obtain a configuration of different S, a ferromagnetic configuration for example; there is a topological barrier stabilizing the skyrmion. Finally, a crucial property of magnetic skyrmion is their solitonic nature: they behave as particles and, for example, can be put into motion by an electrical current. As the skyrmions combine mobility and non-volatility, they are of high interest for new types devices that we describe in more details hereafter.

The specific chirality of the spin configuration in a skyrmion, in most systems, is given by the existence of chiral interactions of the Dzyaloshinskii-Moriya type (DMI) [89], [90].

$$H_{DMI} = (\boldsymbol{S_1} \times \boldsymbol{S_2}) \cdot \boldsymbol{d_{12}} \quad \text{(Eq.2)}$$

Here S1 and S2 are neighbour spins and d12 is the DMI vector. The DMI is a chiral interaction lowering or increasing the energy of the spins depending on whether the rotation from S1 to S2 around d12 is clockwise or counter-clockwise. If S1 and S2 are initially parallel in a ferromagnetic material, the effect of a sufficiently strong DMI interaction is to introduce a relative tilt around d12.

SOC induces such DMI interactions in systems without inversion symmetry. DMIs exist in magnetic compound of non-centrosymmetric crystal lattice and when the breaking of inversion symmetry is due to disorder or to the presence of an interface between different materials. For non-centrosymmetric crystal lattices, the DMIs have been introduced in 1958 and 1960 by Dzyaloshinskii and Moriya as a term of the super-exchange interactions to explain the magnetic properties of insulating magnetic oxides such as $\alpha$-Fe2O3 [89], [90]. In 1980, Fert and Levy [91] showed that DMIs mediated by conduction electrons also exist in magnetic metals and explain some properties of disordered dilute magnetic alloys doped with heavy impurities such as Pt. This picture was then extended to predict the existence of DMIs at interface of magnetic metals [92], as shown in the picture of Fig.9a representing the DMI vector at the interface between a magnetic metal (Fe) and an heavy metal of large SOC (Pt). Today the interfacial DMIs have been derived by ab-initio calculations in many types of interfaces [93], [94]. As shown in Fig.9b, DMIs affect predominantly the spins of the magnetic atoms just at the interface. It decreases rapidly, with possible change of sign, when one moves away from the interface, what gives that a global effect on a magnetic layer is effective for only very thin layers. Several experimental methods are used to measure the interfacial DMI, Brillouin light scattering being the most direct [95], [96].

Skyrmions induced by bulk DMIs were first identified in single crystals with a non-centrosymmetric lattice [97], [98], generally in applied magnetic field and at low temperature. These skyrmions are generally organized in a periodic lattice, as in the image of Fig. 8c taken on the non-centrosymmetric magnetic compound, FeCoSi [99]. Spin textures induced by interfacial DMIs were first observed in the form of spin spirals on ultra-thin films of Fe on W(110) [100]. The first skyrmions induced by interfacial DMIs were identified in the form of skyrmion lattice in a Fe monolayer on Ir(110) [101]. The skyrmions can form a skyrmion lattice but can also exist as non-coupled or weakly coupled individual skyrmions, as shown for the skyrmions represented by Fig. 8d in PdFe bilayer on Ir(111) [86]. Such individual skyrmions, because they can be manipulated individually as nano-balls, are the most interesting for applications. The difference between skyrmion lattices and individual skyrmions is related to the nature of the ground state of the system. If the state with skyrmions is the ground state, this state is a skyrmion lattice. If the ground state is a classical magnetic configuration, ferromagnetic state for example, individual skyrmions appear as metastable states, magnetic defects stabilized by their topological protection [102], [103].

A crucial challenge for applications was the stabilization at room temperature (RT) of small skyrmions. The ordering temperature of bulk materials hosting skyrmions are generally below RT or not much higher than RT. The skyrmions in ultra-thin films (one of a few atomic layers) of the earliest experiments are stable only at low temperature. A direction taken by several groups is the development of multilayers stacking layers of magnetic metals and heavy nonmagnetic metals with additive DMIs at successive interfaces, as in the multilayer in Fig.9c. As the DMIs for Co on Ir and Pt are opposite, the symmetry of the interfacial DMIs makes that the DMIs induced by Ir and Pt in a trilayer Ir/Co/Pt are additive. By repeating these trilayer stacks, the interlayer interactions between successive trilayers couple the skyrmions in a single column, a 3D skyrmion of large magnetic volume of spins which can be stable at RT. Several groups [104]–[107] work today with skyrmions stabilized at RT in this way. Some groups have been able to stabilize skyrmion with diameters as small as 10 nm at RT [108].

The next steps on the way to applications were the development of mechanisms to create skyrmions on purpose, put them into motion and detect them. Fig. 10a shows an example of skyrmion nucleation by the Spin Transfer Torque induced near a constriction by current pulses [109], [110]. Nucleation can also be achieved by replacing constrictions by prepared defects or local injectors from outside the film [103]. Skyrmions in a track can be detected by harnessing the Anomalous Hall Effect of the magnetic multilayer, as illustrate by Fig.10b for the detection of a single skyrmion [111]. Finally, skyrmions can be moved by current-induced Spin Transfer Torques. It turned out that the best mechanism is based on the use of the pure spin current generated by the SHE in the heavy metal layers below or above the magnetic layers in the multilayer [103], [112] (for the generation of vertical spin current by SHE, see the paragraph on SHE and Fig.6d). The velocity of the skyrmions can be very high, up to 100 m/s (at 100m/s, 50 ps is needed to cover 20 nm in a nano-device) in the experimental results in Fig.10c [106].

The interest of skyrmions for technology comes from their properties of quasi-particle, not only ultra-small but also non-volatile. For example, they present all the basic functions needed for information storage and processing: the data can be coded in the form of a train of skyrmions, with writing by nucleation of individual skyrmions at a given position, reading by individual detection of skyrmions at another position, easy current-induced displacement of the information from the nucleation to the detection, logic operations by duplicating or merging skyrmions and easy storage due to the non-volatility.

We will describe only one of the most largely discussed applications of today, the skyrmion-based race track memory. The massive storage of information today, in data centers for example, is based on the use of hard disk drives in which the information is stored by "magnetic bits" fixedly written in the magnetic film covering the disk. Finding a given bit of information is done by rotating the disk under the arm of the write/read head (Fig.3c) and rotating the arm itself. The writing/reading operations thus imply a mechanical system which is complex, fragile and consumer of energy. The race track memory is based on the concept of coding data by a succession of magnetic domain walls in a magnetic track in the first suggestion of Parkin et al [113]. Today with skyrmions, the concept is coding by a train of skyrmions that can be electrically displaced along a track between the points where the skyrmions can be created and detected [114]–[116]. With respect to the hard disk of today, the advantage is that of a purely spintronic and nanoscale device without any mechanical motion of purely electronic devices for a much lower energy consumption. A schematic of skyrmion base race-track memory is shown in Fig.10d. With respect to race-track memory with domain walls, the advantage with skyrmions comes from the flexibility of their shapes and trajectories (reduced pinning), with the possibility of motion in curved tracks (Fig.10d). However, the competition of this technology with 3D Flash memories should be quite severe, especially if this three dimensional integration allows flash memories to reduce their price per bit stored by more than one order of magnitude. Skyrmion-based devices have also been designed to make logic gates in which the duplication or merging of skyrmions is used to perform logic operations [117]. We could also cite the exploitation of the interaction of magnons with skyrmions for the design of skyrmion magnonic crystals for tunable rf filtering [118] or the development of rf oscillators based on the specific dynamical modes of skyrmionic spin configurations [119].

## 6 Conclusion

In this article, we have described a part of the story of spintronics, from its foundation in pure physics of the electronic properties of the ferromagnetic metals to the discovery of GMR, TMR, STT and some very recent developments with topological materials and skyrmions. Unfortunately, in a review of limited length, some other recent developments of spintronics could not be included. Remarkable breakthroughs have also been achieved in spintronics with oxides, a field of research that is generally called oxitronics [50], [51] and includes, for example, the research on the multiferroic materials gathering ferroelectric and ferromagnetic properties. Other interesting directions are spintronics with semiconductors, spin-qbits, spintronics for neuromorphic computing devices, spintronics with graphene and other bidimensional materials as transition metal dichalcogenides, molecular spintronics…On the side of the applications, we have described how spintronics has increased considerably the capacity of data storage and is now bringing its STT-RAM and SOT-RAM components to reduce the energy consumption of our computers and telephones. What will be the next important innovations? The ultra-low power topological logic devices proposed by the researchers at Intel Corporation [84] the STT-oscillator-based neuromorphic devices designed for deep learning [120] ?

# Figures

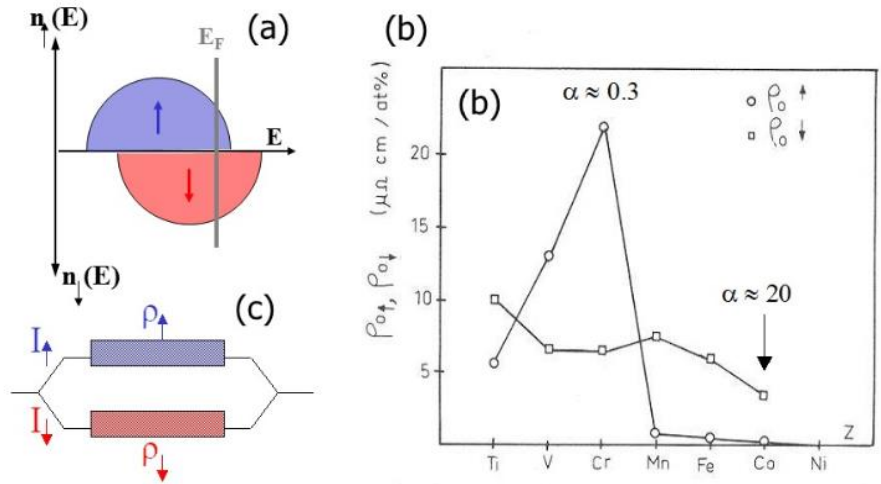

**Figure 1**: Basics of spintronics. (a) Schematic band structure of a ferromagnetic metal. (b) Resistivities of the spin up and spin down conduction channels for nickel doped with 1% of several types of impurity [5] (measurements at 4.2 K). The ratio α between the resistivities $\rho0\downarrow$ and $\rho0\uparrow$ can be as large as 20 (Co impurities) or, as well, smaller than one (Cr or V impurities). (c) Schematic for spin dependent conduction through independent spin up and spin down channels in the limit of negligible spin mixing ($\rho\uparrow\downarrow = 0$ in the formalism of Ref. [5], [6]).

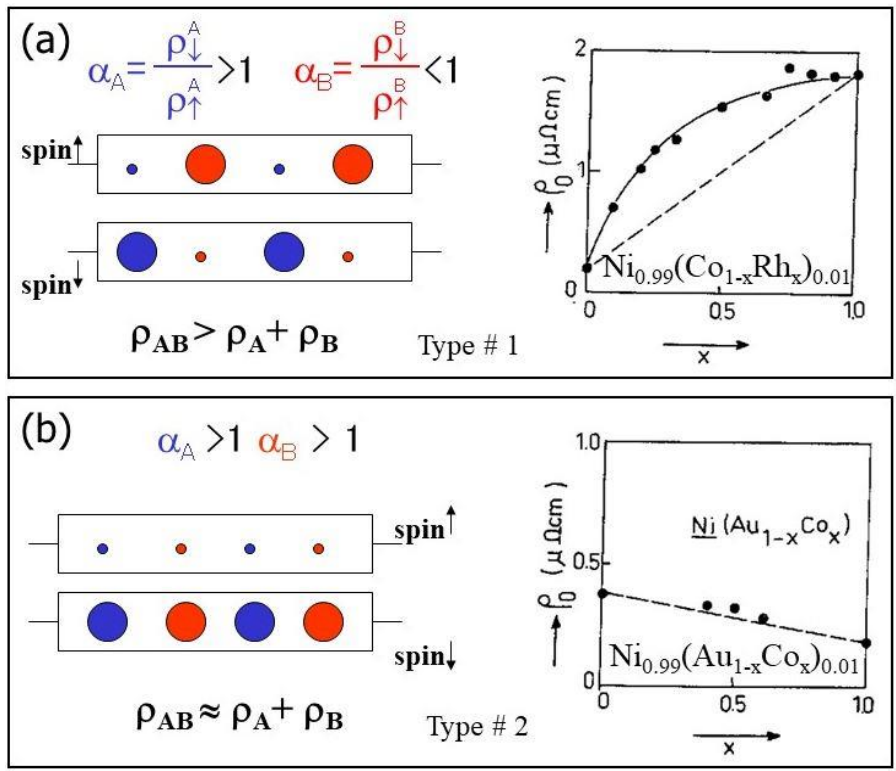

**Figure 2:** Experiments on ternary alloys based on the same concept as that of the GMR. On the sketches illustrating the conduction by two channels in a ferromagnet doped with impurities A (blue) and B (red), the circles are at the scale of the scattering cross sections of impurities A and B. (a) Schematic for the spin dependent conduction in alloys with impurities of opposite scattering spin asymmetries ($\alpha_A = \rho_A\downarrow/\rho_A\uparrow > 1$, $\alpha_B = \rho_B\downarrow/\rho_B\uparrow < 1$, $\rho_{AB} >> \rho_A + \rho_B$) and experimental results on the resistivity of $Ni0.9(Co_{1-x}Rh_x)0.1$ alloys. (b) Same for alloys with impurities of similar scattering spin asymmetries ($\alpha_A = \rho_A\downarrow/\rho_A\uparrow > 1$, $\alpha_B = \rho_B\downarrow/\rho_B\uparrow > 1$, $\rho_{AB} \approx \rho_A + \rho_B$) and experimental results for $Ni_{0.9}(Au_{1-x}Co_x)_{0.1}$ alloys. Figures adapted from [9] (similar results in [4]–[6]). In GMR the impurities A and B are replaced by multilayers, the situation of a (b) corresponding to the antiparallel (parallel) magnetic configurations of adjacent magnetic layers.

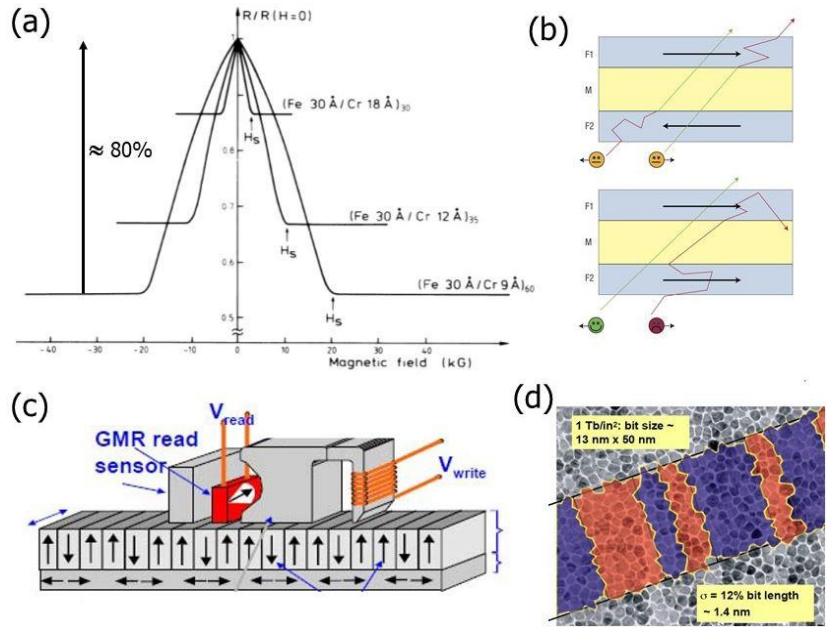

**Figure 3**: Discovery of GMR. (a) GMR in Fe/Cr(001) multilayers [1]. With the current definition of the magnetoresistance ratio, MR= 100x[$R_{AP}$-$R_P$/$R_p$], MR = 85% for the Fe 3nm/Cr 0.9nm multilayers. GMR effects were observed at about the same time with Fe/Cr/Fe trilayers in the team of Peter Grünberg at Jülich [2] (b) Schematic of the mechanism of the GMR. In the parallel magnetic configuration (bottom), the electrons of one of the spin directions can go easily through all the magnetic layers and the short-circuit through this channel leads to a small resistance. In the antiparallel configuration (top), the electrons of both channels are slowed down every second magnetic layer and the resistance is high (figure from Ref. [21]). (c) Schematic of a GMR write/read head in HDD (courtesy of Dr. C. Tiusan). (d) Example of magnetic bits on HDD of today (courtesy of Dr. J. Childress).

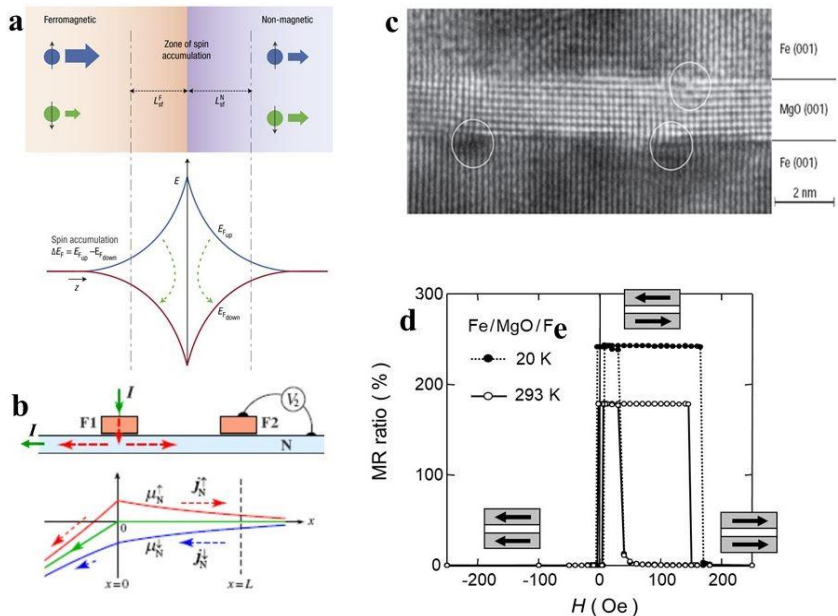

**Figure 4**: (a) Schematic representation of the spin accumulation (out of equilibrium splitting between the spin up and spin down electro-chemical potentials) at an interface between a ferromagnetic metal and a non magnetic layer crossed by a current (arrows symbolize the spin flips induced by the spin accumulation and controlling the progressive depolarization of the current). Figures from Ref. [21]. (b) Schematic of a device in which a current flowing from the ferromagnetic F1 to the left side of the nonmagnetic channel N induces a spin accumulation around the FI/N interface. This spin accumulation generates a pure spin current in the right side of N (zero charge current, only spin flux). (c) Electron microscopy image of a Fe(001)/MgO(001)/Fe(001) MTJ [36]. (d) TMR curves (résistance vs field of a Fe/MgO/Fe MTJ [36].

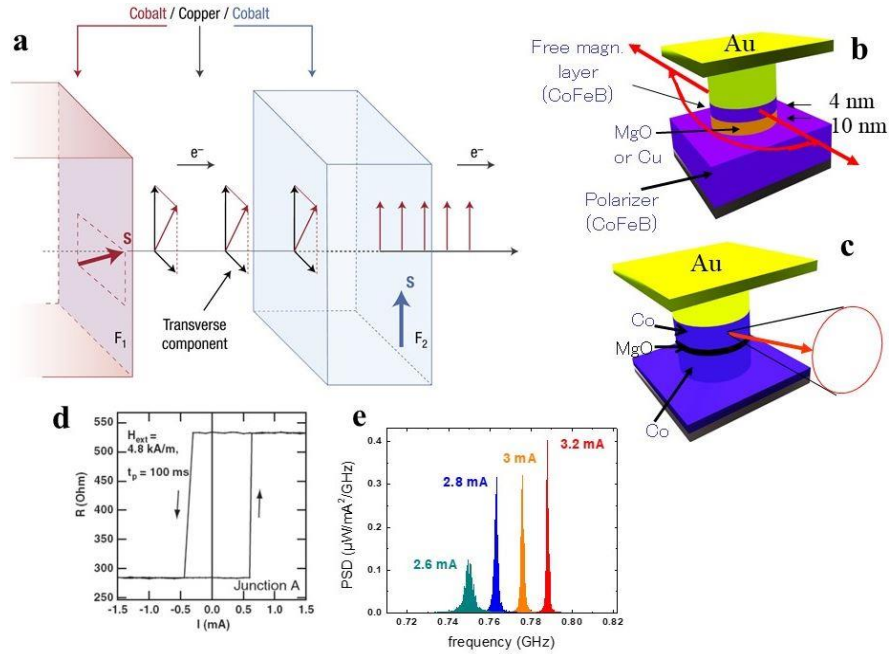

**Figure 5**: (a) Illustration of the spin transfer concept introduced by John Slonczewski [52] and Berger [53]. A spin-polarized current is prepared by a first magnetic layer F with an obliquely oriented spin-polarization with respect to the magnetization axis of a second layer F2. When it goes through F2, the exchange interaction aligns its spin-polarization along the magnetization axis. The exchange interaction being spin conserving, the transverse spin-polarization lost by the current is transferred to the total spin of F2, which can also be described by a spin-transfer torque acting on F2. Figure from Ref. [21]. (b) MTJ nano-pillar (as well, schematic of STT-RAM) for experiments of magnetic switching by STT: the magnetization of the free magnetic layer is reversed by the STT due to injection of a spin-polarized current created by the CoFeB layer. An experimental example of back and forth switching between parallel and antiparallel of MTJ in Fig.5d [121]. (c) Same as (b) but with creation of a steady-states gyration of the magnetization of the free layer (STT oscillator). (e) Example of microwave power emission from gyration of a magnetic vortex in a STT-oscillator.

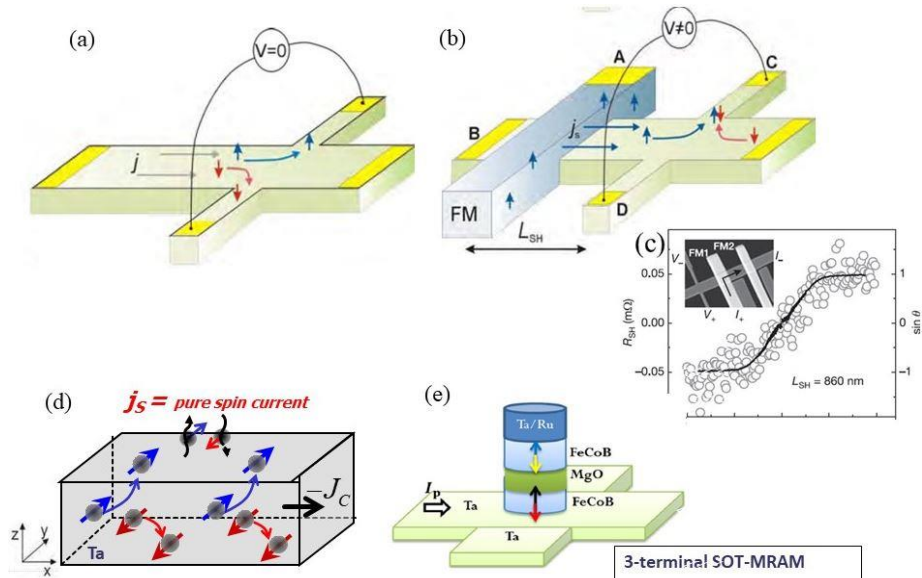

**Figure 6**: (a) SHE: spin accumulation on the sides of a conductor due to the deflection of opposite spins in opposite directions. No charge accumulation and V = 0. (b) Device for the first electrical demonstration of SHE: with opposite spins moving in opposite directions in a situation of pure spin current (as explained for the device of Fig.4b), they are deflected in the same direction into the transvers channel and V ≠ 0. An example of experimental result [66] is in (c). Schematics in (a) and (b) from [67]. (d) SHE with deflection of blue and red spins upward and downward and injection of a pure spin current into an adjacent layer. (e) SOT-RAM: the SHE in the Ta layer injects a spin current into the bottom CoFeB layer to reverse its magnetization and switch the memory.

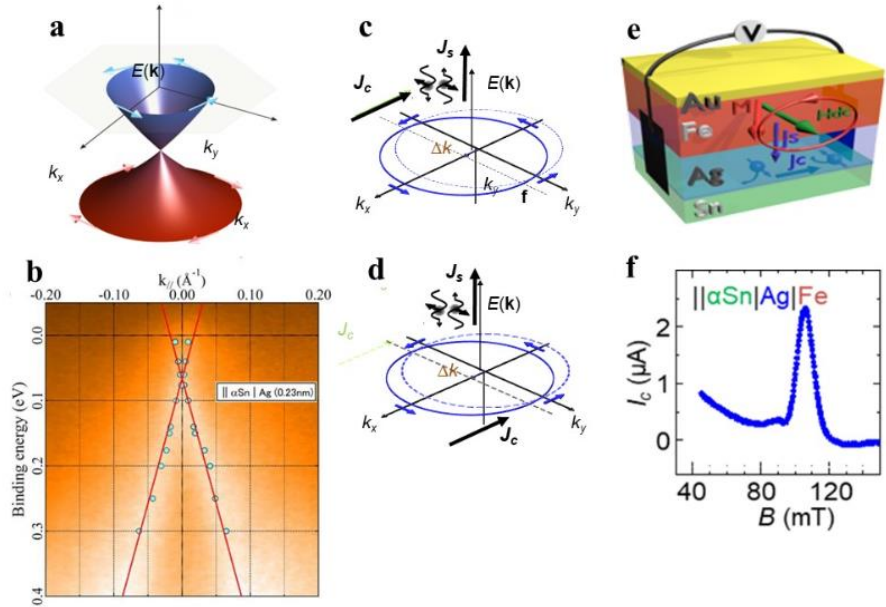

**Figure 7:** (a) Dirac cone of the dispersion surfaces of the 2D electrons at the surface or interface of a topological insulator (TI). Arrow indicate the locking of the spin orientations with the momentum direction. (b) ARPES image of the Dirac cone at the interface of the $\alpha$-Sn TI with Ag [80]. (c) Edelstein Effect (EE): more spins along +y for a current along - x and corresponding shift toward + x of the Fermi contour and production of a vertical pure spin current polarized along +y. (d) Inverse Edelstein Effect (IEE): injection of spin along + y on the + x side of Fermi contour for production of a charge current along x. (e) Spin pumping device for injection of a vertical spin current into the $\alpha$-Sn/Ag interface states by ferromagnetic resonance in the Fe layer and production of a horizontal charge current $I_c$ in these states by IEE. (f) Experimental $I_c$ at the resonance field in the device of (e) [80].

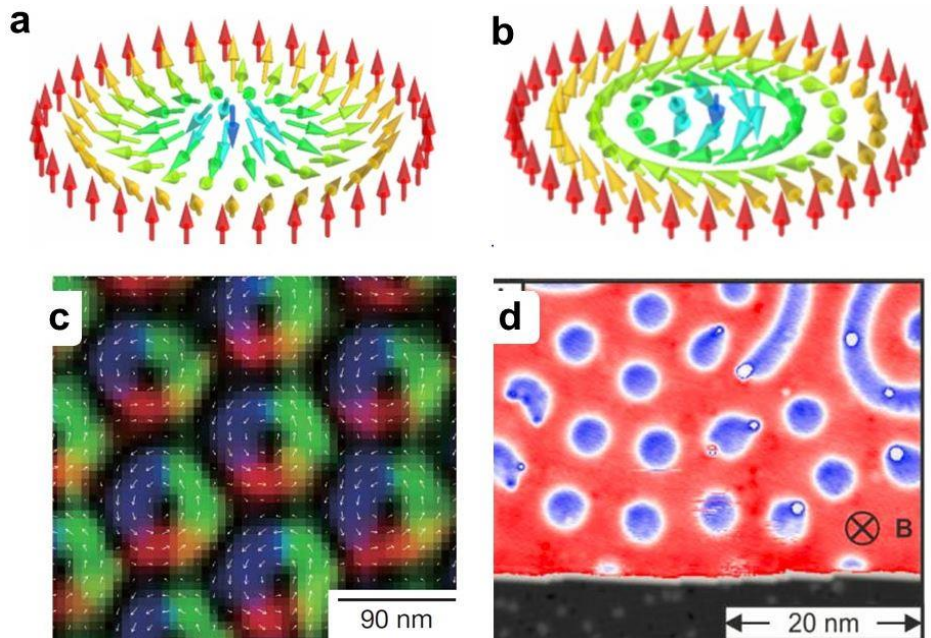

**Figure 8:** (a-b) Spin configurations of Néel (a) and Bloch (b) skyrmions in a magnetic film of out-of-plane magnetization. (c) Lorentz microscopy image of a skyrmion lattice in the non-centrosymmetric magnetic compound Fe0.5Co0.5Si [99]. (d) Spin Polarized Scanning Tunnelling Microscopy image of individual skyrmions induced by interfacial DMI in a Fe/Pd atomic bilayer grown in the heavy metal Ir [122].

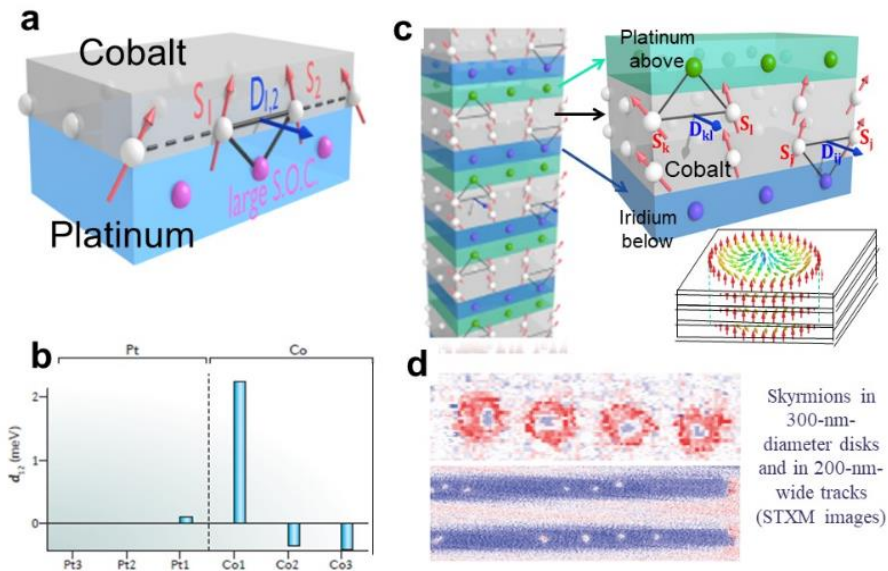

**Figure 9:** (a) Schematic of DMI at the interface between a Co film and the heavy metal P. D12 is the DMI vector in the interface plane. (b) Calculated layer resolved DMI strength in a Co atomic trilayer at interface with Pt [94] (c) Multilayer with additive DMI at the Ir/Co and Co/Pt interfaces. The exchange interactions between successive Co layers couple the skyrmions in a single column of skyrmions [104]. (d) Skyrmions in the multilayer of (c) [104].

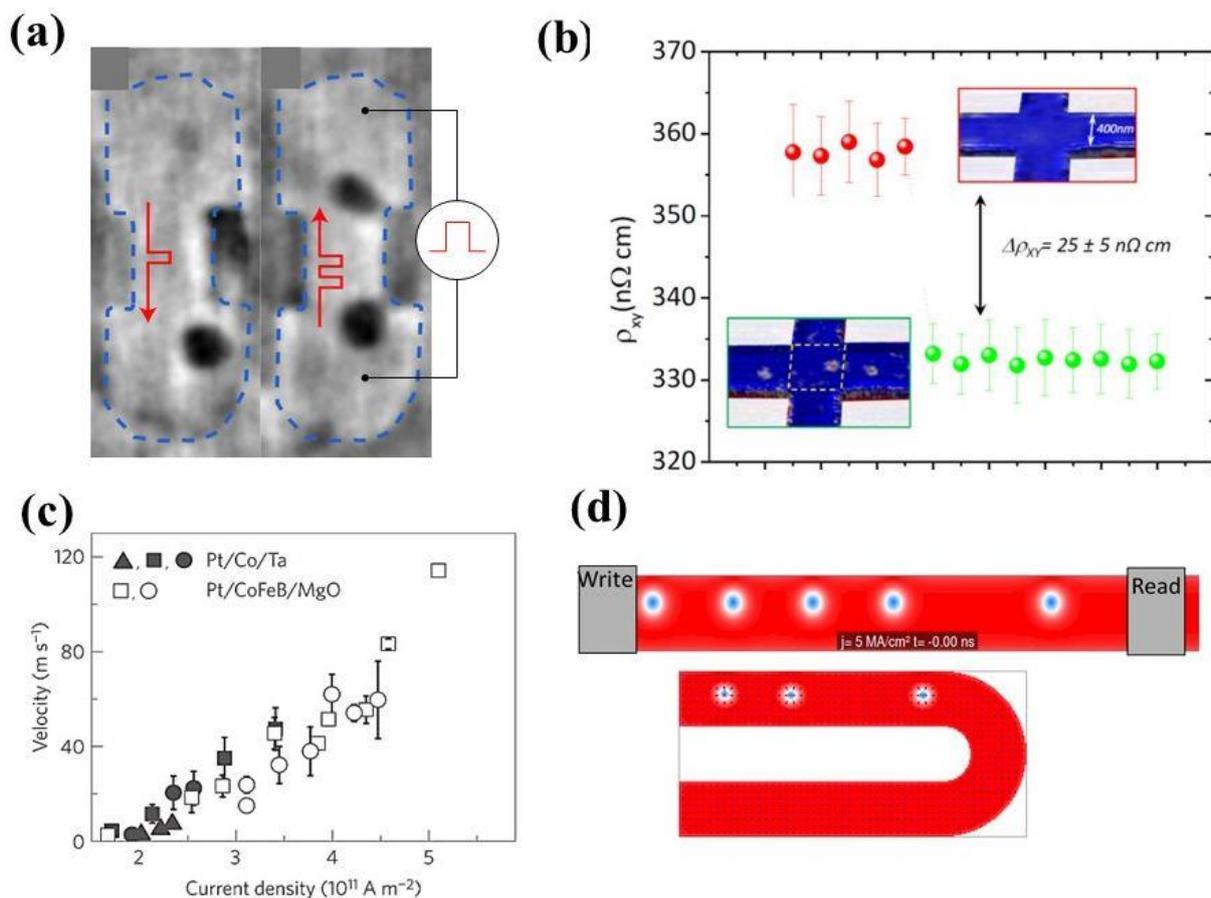

**Figure 10**: (a) Creation of skyrmions at the edge of a constriction in a (Pt/CoFeB/MgO) x 15 multilayer by successive current pulses in opposite directions[110]. (b) Detection of the presence of a single skyrmion by the change in AHE [111]. (c) Skyrmion velocity vs current density in different types of multilayers [106]. (d) Schematic of a skyrmion-based race track memory [103].

**Appendix**

*As Ref [92] (A. Fert, "Magnetic and Transport Properties of Metallic Multilayers," Mater. Sci. Forum, vol. 59–60) is not easily electronically accessible, we present the pages (244-245) describing the mechanism of interfacial DMI. In this proceedings of a 1990 Summer School, the description of interfacial DMI was very short; at this time, the priority was the results on GMR.*

We proceed to the question of the anisotropic pair interactions between moments on surfaces or interfaces. The exchange interaction between two localized moments is generally written:

$$H_{ech} = -J\, \mathbf{S}_1 \cdot \mathbf{S}_2 \qquad (Eq.\ 1)$$

In the most general case, anisotropic interaction terms should also exist, first bilinear antisymmetric terms of the Dzyaloshinskii-Moriya (DM) form [89].

$$H_{DM} = \mathbf{D}_{12} \cdot (\mathbf{S}_1 \times \mathbf{S}_2) \qquad (Eq.\ 2)$$

and higher order terms. The DM term is ruled out when there is a center of inversion symmetry in the system, so that it generally does not exist in ordered crystal lattices, except for some complex structures [89]. However, in disordered systems, in spin glasses for example, there is no longer a center of inversion symmetry, and the DM interactions are known to be significant (~10% of the isotropic exchange, Eq. 1 [91]). The calculation of the DM interaction for the standard geometry of Fig.1 - triangle of Mn impurities in Cu for example – gives a DM interaction, Eq. 1, with $D_{12}$ perpendicular to the plane of the triangle [91].

**Figures**

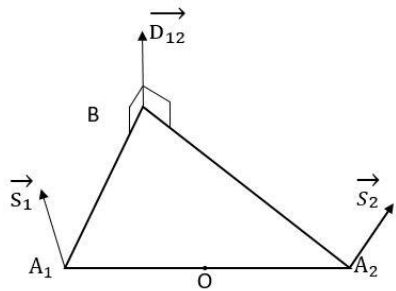
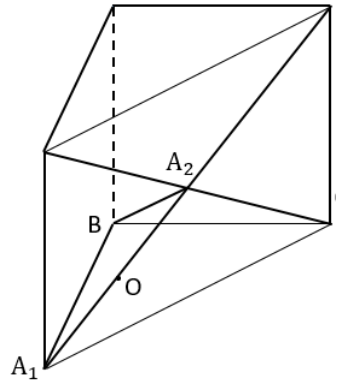

Fig.1          Fig.2

**Figure 1**: Low symmetry geometry giving rise to DM interactions in disordered magnetic alloys. For a triangle of magnetic impurities $A_1A_2B$, the spins $S_1$ and $S_2$ in $A_1$ and $A_2$ are coupled by $H_{DM} = \mathbf{D}_{12} \cdot (\mathbf{S}_1 \times \mathbf{S}_2)$ where $D_{12}$ along $BA_1 \times BA_2$.

**Figure 2**: Low symmetry for a pair of magnetic atoms $A_1$ and $A_2$ on a bcc(110) surface or interface.

In the same way the symmetry breaking at surfaces and interfaces should give rise to DM interactions. For example, in the case of a (110) surface or interface of bcc lattice, see Fig.2, there is no symmetric atom of B with respect to O (as in the absence of surface) and a DM interaction with $D_{12}$ perpendicular to the plane of the triangle $A_1A_2B$ is expected. Such DM interactions are also expected at surfaces or interfaces of fcc and hcp lattices. For both rare-earth and transition metals, one estimates $|D_{12}|/J \approx 10^{-1}$ [91].The effect of such anisotropic interactions on the macroscopic anisotropy properties have never been worked out. For non-S rare-earth ions, it is likely that the effect of the DM interactions is not important with respect to the very strong surface crystal fields. For gadolinium S-ions the crystal field is much smaller and the DM interactions should play a role in the anisotropy properties. In ferromagnetic transition metals, because the exchange is strong, the DM interactions should be significant and likely should contribute to the anisotropy properties. Of course the anisotropic pair interaction effects we have described suppose a localized moment picture of the ferromagnetism and do not turn out in the band structure models usually applied for transition metal ferromagnets.